\documentclass[superscriptaddress,aps,twocolumn]{revtex4}
\usepackage{amssymb}
\usepackage{lineno}
\usepackage{graphicx}
\usepackage{soul}
\usepackage[colorlinks=true,citecolor=blue,linkcolor=blue]{hyperref}
\usepackage[usenames]{color}

\begin{document}

\title{Theory of ball lightning}
\author{H.-C. Wu}
\thanks{huichunwu@zju.edu.cn, http://mypage.zju.edu.cn/en/hcwu}
\affiliation{Institute for Fusion Theory and Simulation (IFTS) and Department
of Physics, Zhejiang University, Hangzhou 310027, China}
\affiliation{IFSA Collaborative Innovation Center, Shanghai Jiao Tong University,
Shanghai 200240, China}
\date{\today}

\begin{abstract}
We present a comprehensive theory on the formation of ball lightning,
a luminous sphere sometimes observed after normal lightning. In a
ball lightning event, a relativistic electron bunch can be produced
by the stepped leader of lightning and coherently emit high-power
microwave when striking the ground. The intense microwave ionizes
the local air and evacuates the resulting plasma by its radiation
pressure, thereby forming a spherical plasma cavity that traps the
microwave. The theory can explain observed properties of
ball lightning and should be useful to lightning protection and aviation
safety.
\end{abstract}

\maketitle

Since Arago \cite{Arago} first discussed ball lightning in 1838,
this rare natural phenomenon still remains a riddle. Ball lightning
\cite{Singer,Barry,Stenhoff,Uman} exhibits very diverse characteristics,
such as close association with ordinary lightning, globate structure
with steady glow for 1-5 seconds, and mostly horizontal motion. Ball
lightning can be formed even inside aircraft and closed rooms, permeate
glass plates, decay explosively or silently, and produce sound and
acrid odours. Many models of ball lightning have been proposed, but
none have been fully accepted \cite{Ball}. In particular, these theories
do not explain the inscrutable appearance of ball lightning inside
fully-screened aircraft \cite{Dijkhuis}. Here, we propose a theory
for ball lightning formation, which can explain its appearing in aircraft
and many other properties.

Lodge \cite{Lodge} considered that ball lightning might be excited
by a standing electrical wave from lightning. Kapitza \cite{Kapitza}
argued that ball lightning could be formed through air ionization
at antinodes of electromagnetic standing waves in the microwave regime.
Dawson and Jones \cite{Dawson} proposed that ball lightning could
be a microwave bubble confined inside a globate plasma shell. Continuous
air ionization by the trapped microwave maintains the plasma shell
\cite{Zheng}. The microwave type model of ball lightning can explain
its permeation through glass plates. However, the origin of microwave
emission from lightning was never found. On the other hand, high-intensity
lasers were observed to get trapped in a spherical plasma cavity in
particle-in-cell (PIC) simulations \cite{Naumova} and experiments
\cite{Sarri,Sylla}. By dimensional analysis, we already pointed out
\cite{Wu1} that a microwave bubble can be formed by the similar mechanism.
A sketch of the microwave bubble model is displayed in Fig. 1a.

Here we propose a mechanism for microwave generation from lightning.
As shown in Fig. 1b, we assume that in a ball lightning event a relativistic
electron bunch is generated by lightning. When this bunch strikes
the ground or passes various media, powerful microwaves are emitted
by coherent transition radiation (Fig. 1c). We further verify that
this specific microwave in air plasmas indeed naturally evolves into
a microwave bubble. These results are demonstrated by PIC simulation
using the code JPIC \cite{Wu1}.

\textbf{Relativistic electron bunch}

The assumption of isolated relativistic electron bunches in ball lightning
events is based on high-energy phenomena \cite{Dwyer1,Dwyer2} discovered
in cloud-to-ground lightning. A lightning flash \cite{Uman} starts
with a negative leader propagating downward in a stepping process
with each step tens of metres. This stepped leader has a corona 1-10m
in width. Moore \textit{et al.} \cite{Moore} first detected $>$1MeV
radiation from a stepped leader. It was then observed that each step
emits an x-ray burst \cite{Dwyer3}, which intensifies when the leader
approaches the ground. Recent data \cite{Howard} shows that the last
step or the so-called leader burst closest to the ground produces
the strongest x-rays. Relativistic electrons accelerated by the stepped
leader account for the detected x-rays, and therefore the electron
acceleration is the most violent in the last step.

Friction force of electron motion in air is maximum at an energy of
100eV, which defines a critical electric field $E_{c}\approx30$MV/m
\cite{Dwyer2}. Fields above $E_{c}$ at the leader tip can accelerate
thermal electrons to several keV \cite{Moss}. This thermal runaway
process \cite{Gurevich1} produces $\sim$10$^{11}$ electrons. These
hot electrons can be further accelerated by the electric field between
the leader tip and the ground, and undergo avalanche by secondary
electron generation in air \cite{Gurevich2}. The electron flux quickly
rises as $\exp(z/L)$, where $L$ is the avalanche length. The electron
energy follows an $\exp(-k_{e}/7.3$MeV$)$ distribution, such that
the average energy is 7.3MeV. Each seed electron multiplies to $\sim$10$^{5}$
electrons after the avalanche, so that each leader step can produce
$10^{16}-10^{17}$ relativistic electrons \cite{Dwyer1} with a total
energy of 10-100kJ.

Accordingly, it can be expected that the last leader step generates
a spatially well-defined relativistic electron bunch in a ball lightning
event. For simplicity, we assume that this bunch has a density profile
$n_{b}=n_{b0}\exp(-r^{2}/2\sigma^{2})$, where $n_{b0}$ is the peak
density, and $\sigma$ is the characteristic radius. Such a bunch
with total electron number $N_{b}=(2\pi)^{3/2}n_{b0}\sigma^{3}\approx10^{14}$
will lead to a microwave bubble. Since this bunch contains only 0.1-1\%
of the generated electrons, the assumption is reasonable.

\begin{figure}[t]
\includegraphics[width=.43\textwidth]{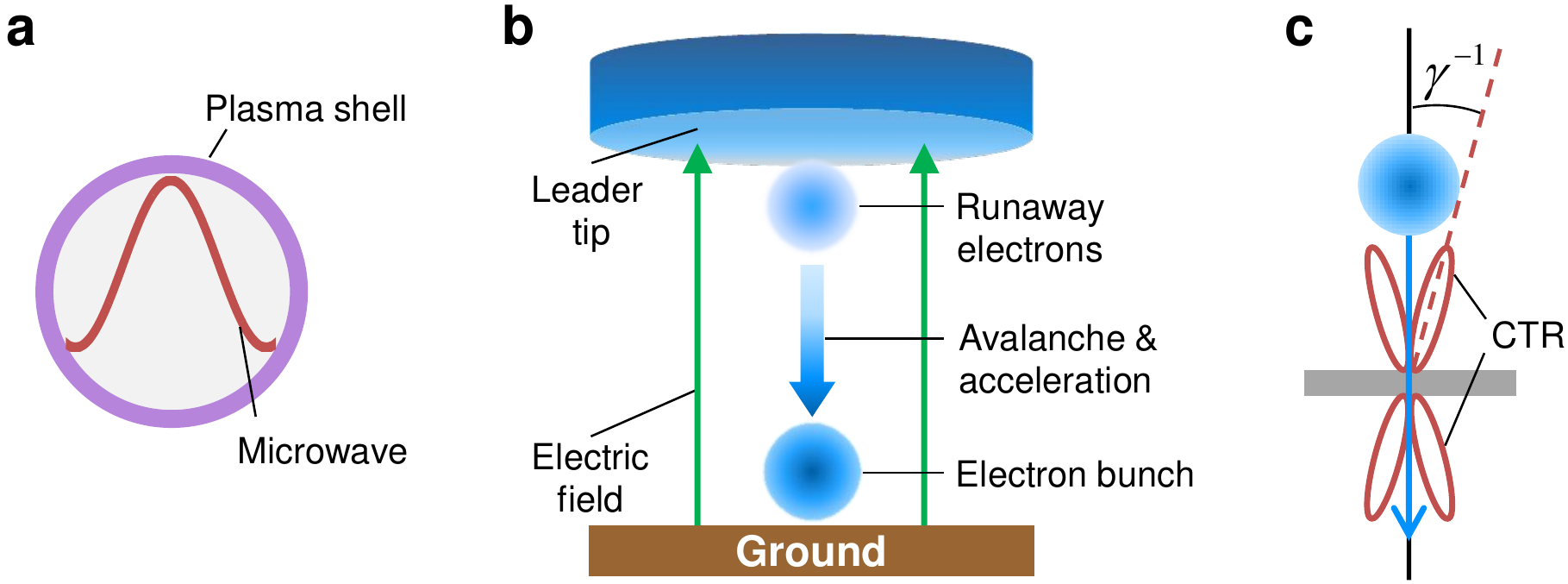} \label{fig1}
\caption{\textbf{Ball lightning model.} a, Microwave bubble model. b,
Relativistic electron bunch generation. In the last leader step, a
bunch of runaway electrons emerges from the leader tip, accelerates
by electric fields between the leader and ground, and undergoes an
avalanche. c, Coherent transition radiation (CTR) of the electron
bunch striking the ground or passing through aircraft skins. $\gamma$
is the relativistic factor of electrons.}
\end{figure}

\textbf{Microwave generation}

Transition radiation is generated from medium surfaces when an electron
enters or emerges \cite{Landau} and can be coherent for an isolated
electron bunch \cite{Happek}. As the electron bunch reaches relativistic
energies, its self-fields are predominantly transverse i.e. $\mathbf{E}_{b}\simeq c\mathbf{B}_{b}$
\cite{Jackson}, which is very close to an electromagnetic wave. In
this case, coherent transition radiation can be considered as the
reflected wave of the bunch field from the medium surface \cite{Casalbuoni}.
Therefore, we can write the radiation energy as
\begin{equation}
W_{CTR}=\mathcal{R}(\varepsilon)W_{b,f},
\end{equation}
where $\mathcal{R}=\left\vert (\sqrt{\varepsilon}-1)/(\sqrt{\varepsilon}+1)\right\vert ^{2}$
is the Fresnel reflection formula, $W_{b,f}$ refers to the total
bunch field energy, and $\varepsilon$ is the medium permittivity.
The radiation is strongest for a metal or perfect conductor where
$\varepsilon\rightarrow\infty$ and $\mathcal{R}\approx1$ in microwave
region.

The leftmost panel of Fig. 2 shows the transverse field $E_{b,x}$
of a monoenergetic 7MeV electron bunch with $\sigma=4$cm, which is
normalized to the peak field $E_{b0}\approx3.2\frac{n_{b0}}{10^{10}cm^{-3}}$MV/m.
The bunch field is a unipolar wave with the same profile as the bunch
density. Using JPIC \cite{Wu1}, we simulate the coherent transition
radiation from a perfect conductor in Fig. 2. The radiation field
$E_{x}$ is initially opposite to $E_{b,x}$ due to the conductor
boundary, diffracts transversely, and quickly evolves into a bipolar
pulse. This radiation has a central wavelength $\lambda\approx7.5\sigma=30$cm
(i.e. 1GHz). The rapid field evolution into the bipolar shape is due
to diffraction losses of longer wavelength components in an unipolar
pulse \cite{Wu2}. At normal incidence in Fig. 2, the radiation field
has a ring-like intensity distribution. In reality, oblique incidence,
surface fluctuations, and non-symmetric bunches will lead to a complex
field pattern containing linearly-polarized emission spots.

\begin{figure}[t]
\includegraphics[width=.35\textwidth]{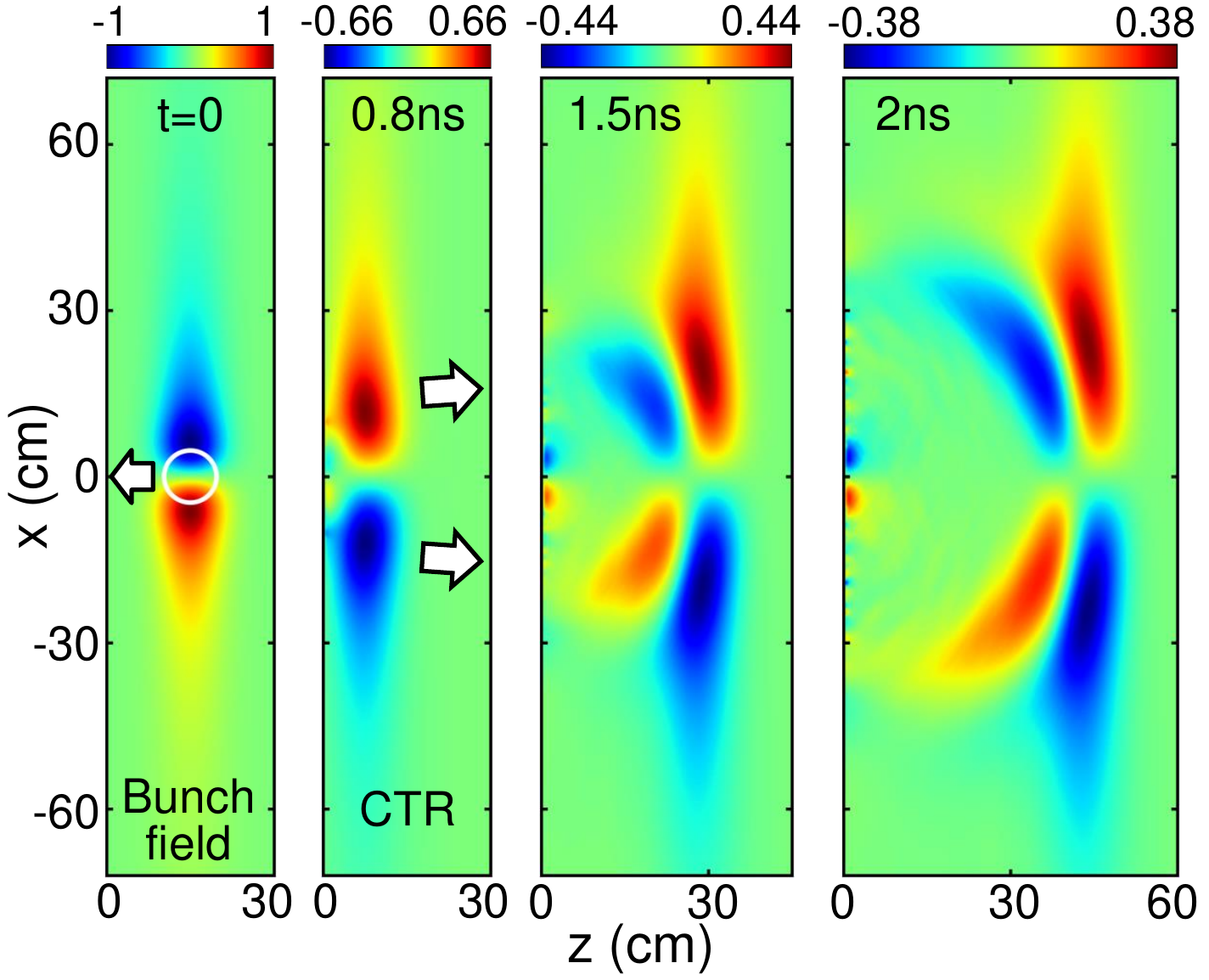} \label{fig2}
\caption{\textbf{PIC results of microwave generation.} Distribution
of the initial bunch field and microwave fields at times 0.8ns, 1.5ns and 2ns. In the leftmost panel, the bunch is left-going to the
plasma surface at $z=0$. The white circle marks the bunch region
with a density of $0.5n_{b0}$. The radiation is a reflection of the
bunch field and propagates along $z$. Arrows point along the field
propagation direction. The field is normalized to the bunch peak field
$E_{b0}$.}
\end{figure}

\textbf{Microwave bubble formation}

Laser bubbles or solitons \cite{Naumova} are formed after tens of
light cycles of interaction between multi-cycle relativistic laser
and underdense plasmas with an initial density $n_{0}<n_{c}$, where
$n_{c}=\varepsilon_{0}m\omega^{2}/e^{2}$ is the critical density
\cite{Kruer}. In our case, the microwave is mono-cycle. Therefore,
the microwave must get trapped within a few cycles before it is diffracted
away. We indeed find such a regime for bubble formation, which indicates
self-consistency of our theory. The collisional effect is included
by embedding air friction \cite{Dwyer2,Moss} into JPIC. We launch
microwave pulses with a wavelength $\lambda=30$ cm into a uniform
plasma. The initial plasma must be overdense with $n_{0}\geq n_{c}$,
where $n_{c}\approx1.2\times10^{10}$cm$^{-3}$ at $\omega/2\pi=1$GHz.
The threshold field required for bubble formation is
\begin{equation}
E_{bl}\approx\max(4E_{c},E_{r}),
\end{equation}
where $E_{r}=mc\omega/e$ is the relativistic field threshold \cite{Wu1}.
At 1GHz, we have $E_{r}\approx10.7$MV/m and $E_{bl}\approx11E_{r}\approx120$MV/m,
which is highly relativistic. Equation (2) clearly shows that the
field needs to be greater than $E_{c}$ to efficiently accelerate
electrons, and reach the relativistic regime to completely expel electrons
by a relativistic pondermotive force \cite{Sun}. Surprisingly, $E_{r}$
matches with $E_{c}$ to make the bubble formation possible.

In Fig. 3, we take $n_{0}=4n_{c}$ and a microwave field of 310MV/m,
and let $t=0$ when the field touches the plasma. The radiation pressure
first pushes electrons to pile up into a semicircular shell and leaves
a low-density region at the rear. As the field is reflected by the
front shell, peripheric electrons return to the low-density region
and close up the cavity. The field gets trapped and then evolves into
a standing-wave mode. At $t\approx8$ns, a stable bubble forms about
40cm deep into the plasma, and it is circular and motionless. Heavy
ions are slowly pulled out by the charge separation field.

In Figs. 3a and 3b, snapshots at $t=18$ns show that the fields take
on a half-cycle standing wave pattern, electrons are almost emptied,
and ions are partially evacuated. The electrostatic force between
electrons and ions is balanced by the radiation pressure $\varepsilon_{0}E^{2}/4\approx$64kPa,
where $E=$170 MV/m is the standing wave amplitude. The periodic conversion
between electric and magnetic energies in Fig. 3c confirms the standing
wave mode. The confined field oscillates with a longer period of 1.6ns.
This redshift is caused by the Doppler effect and self-phase modulation
\cite{Watts}. The cavity diameter is about $24$cm, half of the wavelength
of the trapped field.

For a ball shape, the confined field energy in Fig. 3b is about 800J.
Tuning the microwave field, the trapped field energy ranges from 200J
to 1500J in the bubble. Finally, we check the bunch parameters for
giving the threshold field $E_{bl}$. For the case in Fig. 2, we get
$n_{b0}\approx3.7\times10^{11}$cm$^{-3}$ and $N_{b}\approx3.7\times10^{14}$.

\begin{figure}[t]
\includegraphics[width=.43\textwidth]{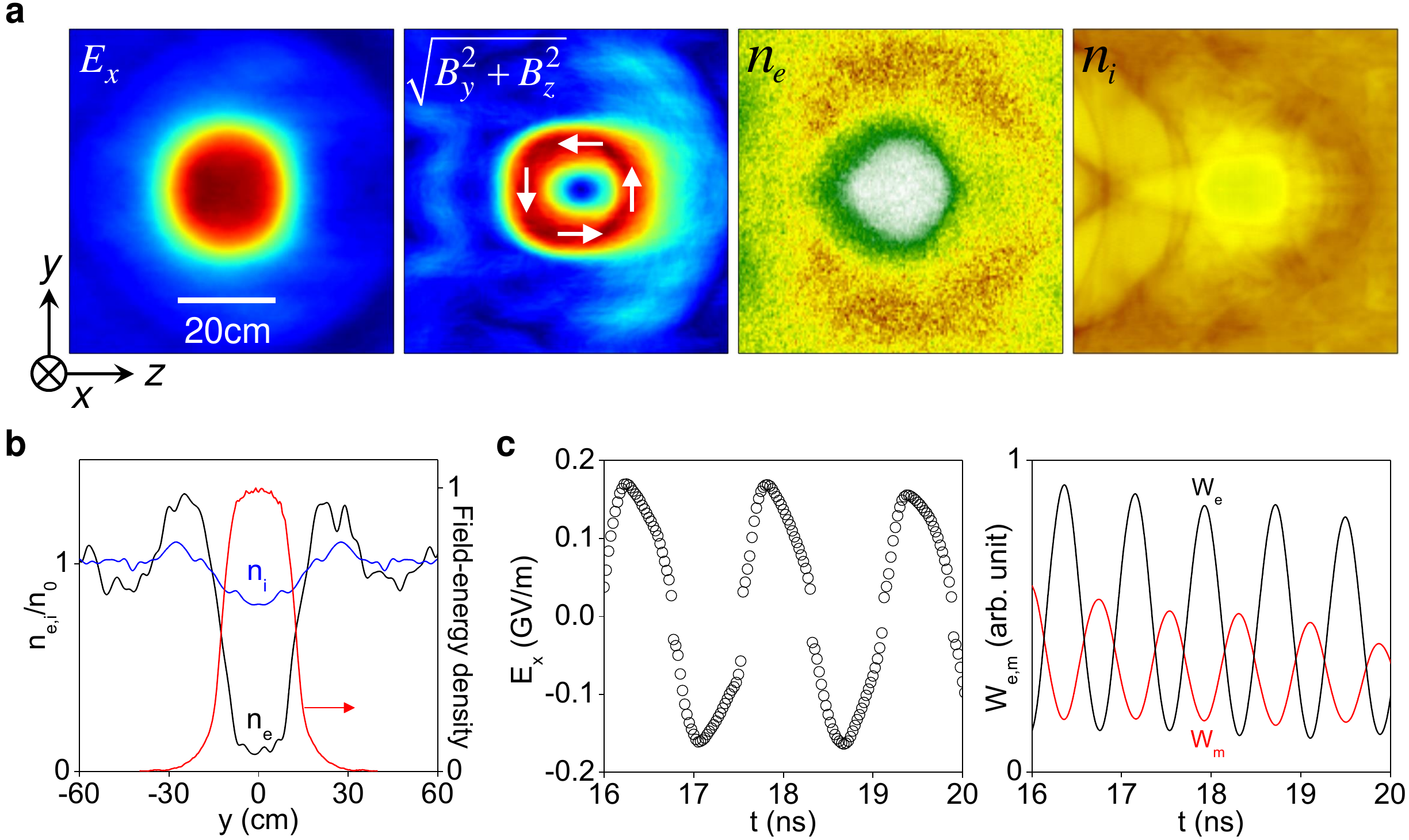} \label{fig3}
\caption{\textbf{PIC results of microwave bubble formation.} a, Snapshots
of the electric field $E_{x}$, magnetic field $\sqrt{B_{y}^{2}+B_{z}^{2}}$,
electron density $n_{e}$, and ion density $n_{i}$ of the microwave
bubble at $t=18$ ns. White arrows mark the magnetic field direction.
b, Field energy density and plasma density verses $y$ across the
bubble centre. c, Evolution of the electric field, electric field
energy $W_{e}$, and magnetic field energy $W_{m}$ in the bubble.}
\end{figure}

\textbf{Explanation of the diverse properties}

The properties of ball lightning \cite{Singer,Barry,Stenhoff,Uman}
are summarized from about 5000 published reports.

\textit{Site of occurrence.} As shown in Fig. 2, a planar surface
is necessary for microwave generation at least with a size of ball
lightning, which can be easily fulfilled in reality. Microwave emission
is also affected by the ground reflectivity $\mathcal{R}$. The soil
permittivity $\varepsilon$ increases with its moisture $m_{s}$ \cite{Hallikainen}.
At 1GHz, we get $\left.\varepsilon\right\vert _{m_{s}=20\%}\approx8.4-2.2i$
and $\left.\varepsilon\right\vert _{m_{s}=60\%}\approx48-10i$, which
correspond to $\mathcal{R}\approx25\%$ and $56\%$, respectively.
Rainfall can lead to $m_{s}>60\%$ \cite{Xu} and thus is favorable
for the ball lightning formation. As stated by Stenhoff \cite{Stenhoff},
more than 50\% of reports show that medium or heavy rainfall happens
before the observation. Moreover, there is $\mathcal{R}\approx65\%$
for either pure or sea water \cite{Meissner}. Indeed, there are 18
reports at sea \cite{Singer} and a few reports over rivers \cite{Singer,Stenhoff}.
Certainly, metal holds the highest chance of ball formation due to
$\mathcal{R}\approx1$.

\textit{Relation to lightning channels.} The lightning channel refers
to the bright return stroke occurring after the stepped leader attaches
with a positive leader rising from the ground. The starting place
of this positive leader would be the lightning strike point. We show
that ball lightning is caused by the stepped leader, which is invisible
with the naked eye. The stepped leader and its mirror charge underground
establish a dark channel for electron acceleration and avalanche.
Obviously, the ball formation site is unrelated to the lightning strike
point. Their separation should be within one step length of tens of
meters typically. This successfully explains the reports where ball
lightning does not form near the lightning channel or strike point
\cite{Stenhoff}.

\textit{Appearance in aircraft.} First, the electron energy 7.3MeV
is independent of the air density, i.e. altitude \cite{Dwyer1}. When
lightning strikes an aircraft, the same bunch is presumably produced
and enters the aircraft with an energy loss of $\sim$2MeV due to
the $\sim$0.6cm aluminium skin \cite{Dwyer4}. Second, transition
radiation \cite{Landau} is not sensitive to the energy of the relativistic
electrons, and its efficiency from the electron emerging surface of
the medium is almost the same as the reflection side discussed above.
Therefore, the same intense microwave will arise inside the aircraft
and form ball lightning there. In the same manner, ball lightning
can appear in enclosed rooms.

\textit{Permeation through glass plates.} Ball lightning is observed
to enter rooms by passing through closed glass windows. In interference
experiments of low-power microwave in metal cavity \cite{Ohtsuki},
generated fireballs in air are observed to pass through a 3mm ceramic
plate intact. This is a direct result of the ability of microwave
passage across dielectrics. The microwave bubble resembles a laser
cavity. According to laser theory \cite{Siegman}, the internal standing
wave will not be disturbed if a glass plate ($\sim$5mm) is much thinner
than the wavelength of microwave.

\textit{Shape.} By dimensional analysis \cite{Wu1}, the microwave
bubble of Fig. 3 in reality should be ball-shaped as its counterpart
in laser experiments \cite{Sarri,Sylla}. The full trapping of the
field in Fig. 2 can account for the 62 ring-shaped ball lightning
reports \cite{Singer}.

\textit{Size.} Ball lightning has a common diameter of 20-50cm \cite{Stenhoff}.
Our theory shows that the diameter of microwave bubbles approximately
equals the bunch length in the direction of motion. The avalanche
length $L$ is 7-30cm near the ground \cite{Dwyer2}, which should
support the generation of tens of cm long bunches.

\textit{Sound.} Hissing, buzzing or fluttering sounds from ball lightning
have been reported, which can be perfectly explained by the microwave
hearing effect \cite{Frey,Lin}. At 0.1mJ/cm$^{2}$, a microwave pulse
(microsecond or shorter) at 0.2-3GHz can induce an audible sound wave.
The sound can only be heard by persons whose heads are irradiated
by the microwave, and has been described as a hiss, buzz or knocking.
Therefore, ball lightning can be silent during its lifetime. In Jennison's
sighting \cite{Jennison}, he was only 0.5m from a cruising ball,
and did not report any noise.

\textit{Spark.} Ball lightning sometimes emits sparks, which can be
caused by the ejection of charged particles along the electric field.
Especially, the sparks are toward opposite directions in two reports
\cite{Singer}, which agrees with the linear polarization of standing
wave in the bubble.

\textit{Spectrum}. Recently, Cen \textit{et al.} \cite{Cen} recorded
an optical spectrum of ball lightning. The spectrum contains emission
lines of atoms in air and soil. Interestingly, the spectral intensities
of O and N atoms oscillate at 100Hz, twice the frequency of the adjacent
power lines (35kV, 50Hz). The latter is only 20m from the ball and
can produce a 50Hz electric field of $\sim$1V/cm \cite{Tzinevrakis}
at the ball. This field can induce electron drift on the ball surface
with an amplitude about tens of cm. This drift motion can perturb
the spectral emission in the plasma shell. Then, the spectral intensity
should be independent of drift direction and varies at 100Hz. The
ball is attached to the soil on a hillside, where electrons cannot
feel the oscillating field due to the screening effect. Thus, Si, Fe and Ca in soil glow steadily
\cite{Cen}.

\textit{Odour}. Ionized air can produce O$_{3}$ and NO$_{2}$ \cite{Uman,Petit},
both of which have an acrid smell.

\textit{Decay}. The microwave bubble decays silently once the internal
radiation is exhausted. When it is strongly disturbed or pierced by
a conductor, the leaking radiation can launch a shock wave like an
explosion \cite{Zheng}.

\textit{Injury and damage.} Most reported injuries and damages can
readily be attributed to ordinary lightning \cite{Singer,Stenhoff}.
However, Stenhoff \cite{Stenhoff} noticed that some superficial burns
are difficult to explain. In the Smethwick event \cite{Stenhoff},
the female witness did not get an electric shock but felt a burning
heat all over. Wooding \cite{Wooding} estimated that she received
250J whole-body ionizing radiation, which can be due to the electrons
from the stepped leader and also be responsible for the redness on
her hand and legs. She heard a sound of knocking-like rattle from the microwave hearing effect.
Her legs were numbed, which can be due to nerve damage by the microwave
at 0.1J/cm$^{2}$ \cite{Benford}. When she brushed the ball away
with her hand, the ring was burning into her finger. Wooding calculated
that this rapid heating would need a resonant microwave at 1GHz with
an field of $\sim$1MV/m, which agrees well with our model. Others \cite{Shmatov}
reported skin redness, vomiting and loss of hair,
which are typical results of ionizing radiation \cite{Mettler}. In
the Shanxi event \cite{Shanxi}, computers are damaged by an exploded
ball indoors as a direct result of intense microwave radiation \cite{Benford}.

\textit{Motion.} Near the ground, ball lightning moves mostly horizontally
at about 2m/s \cite{Singer} and usually travels with the wind \cite{Barry}.
A light breeze typically at 1.5-3m/s \cite{Beaufort} can account
for this motion speed. However, air convection will raise the ball
if the background air is heated up by the ionized plasmas. Assuming
a constant heat power of 100W, we obtain a convection speed 23cm/s
for the ball of size 30cm. Thus, the upward motion is not notable
compared with the horizontal motion. Several models \cite{Singer,Stenhoff}
speculate that the ball could take a positive charge due to the greater
mobility of electrons compared with ions. The charged ball can further
resist the buoyancy or air convection by an attractive force from
its mirror charge underground. Moreover, like a charged particle self-accelerating
into an open waveguide \cite{Tyukhtin}, the ball can enter rooms
through chimneys.

\textit{Lifetime.} The typical lifetime of ball lightning is 1-5 seconds.
Statistical analysis \cite{Amirov} shows that increase in humidity
decreases the lifetime of the ball, which can be due to microwave
absorption by vapour. Experiments \cite{Ofuruton} show that fireballs
in air produced by a 5kW, 2.45GHz microwave can last for $\sim$0.5s
after the source is turned off. Our self-organized microwave bubble
can have the same potential to persist for a scale of seconds. Zheng
\cite{Zheng} quantitatively calculated that hundreds of joule microwaves
can maintain the plasma shell of the bubble for a few seconds. Air
plasmas continuously depleted by recombination are refilled by microwave
heating. Non-neutral plasmas shown in
Fig. 3b can further resists the recombination loss.

We need to obtain a more convincing depletion rate of the microwave,
find a balance among the radiation pressure, electrostatic force of
the non-neutral plasmas, and air resistance or pressure felt by the
plasma shell, calculate the luminosity and net charge of the bubble,
and check whether the bubble can keep the constant appearance throughout
its lifespan. These studies rely on a precise knowledge of the plasma
shell, which should consider the ionizing process of air and chemical
reactions \cite{Moss} in air plasmas.

\textbf{Experimental suggestions and conclusions}

Experiments are required to verify our theory. First, forming a microwave
bubble in laboratory will need hundreds of gigawatt microwave, which
is one order of magnitude higher than the manmade sources. As stated
in Ref. \cite{Benford}, it is technically feasible to enhance current
microwave devices to 100GW. Alternatively, one can adopt a high-power
electron beam \cite{Humphries} to directly simulate the mechanism
proposed in Fig. 1. Second, on the lightning research, we suggest
to detect microwave radiation at GHz near a lightning strike point.
We already show that trans-ionospheric and sub-ionospheric pulse pairs
from lightning are caused by the same radiation mechanism \cite{Wu3} (or see http://arxiv.org/abs/1411.4784v1),
which supplies a physical evidence of our theory. On attempts to create
ball lightning by rocket-triggered lightning \cite{Hill}, we propose
to use ungrounded wires \cite{Uman} rather than grounded ones because
ball lightning is thought to be only related to the stepped leader.
Finally, for in situ investigation of ball lightning, we suggest to
look for evidence of high-flux energetic electrons.

In conclusion, starting with a reasonable assumption on the electron
bunch, we construct a self-consistent theory on the microwave generation
and ball lightning formation. The theory successfully explains many
properties of ball lightning. For the first time, we revel that ball
lightning is an alarm signal of the existence of ultrastrong microwaves
and abundantly hazardous electrons near the ground or aircraft. This
result is of great significance for lightning protection and
aviation safety. Moreover, it is hoped that our work will stimulate
research activities in relativistic microwave physics and technology,
an unexplored area.

\begin{acknowledgements} This work was supported by the Thousand
Youth Talents Plan, NSFC (No. 11374262), and Fundamental Research
Funds for the Central Universities. We thank W. M. Wang and S. M.
Weng for discussions on collisional PIC, Z. H. Wang for instruction
in laser cavities, and M. Y. Yu for helpful comments. \end{acknowledgements}

\end{document}